\documentclass[12pt]{article}
\usepackage{floatrow}
\usepackage{times}
\usepackage{geometry}
\usepackage[utf8]{inputenc}
\usepackage{enumitem,amssymb}
\usepackage{ragged2e}
\newlist{thematic}{itemize}{8}
\usepackage{hyperref}
\setlist[thematic]{label=$\square$}
\usepackage{pifont}

\usepackage{graphics,graphicx} 
\usepackage[font={small,it}]{caption}
\begin{document}
\raggedright
\huge
Astro2020 Science White Paper \linebreak

A Unification of the Micro and Macro Physics in the Intracluster Medium of Nearby Clusters \linebreak
\normalsize

\noindent \textbf{Thematic Areas:} \hspace*{60pt} $\square$ Planetary Systems \hspace*{10pt} $\square$ Star and Planet Formation \hspace*{20pt}\linebreak
$\square$ Formation and Evolution of Compact Objects \hspace*{31pt} \makebox[0pt][l]{$\square$}\raisebox{.15ex}{\hspace{0.1em}$\checkmark$} Cosmology and Fundamental Physics \linebreak
  $\square$  Stars and Stellar Evolution \hspace*{1pt} $\square$ Resolved Stellar Populations and their Environments \hspace*{40pt} \linebreak
  \makebox[0pt][l]{$\square$}\raisebox{.15ex}{\hspace{0.1em}$\checkmark$} Galaxy Evolution   \hspace*{45pt} $\square$             Multi-Messenger Astronomy and Astrophysics \hspace*{65pt} \linebreak

\textbf{Principal Author:}

Name: Yuanyuan Su	
 \linebreak						
Institution: University of Kentucky 
 \linebreak
Email: ysu262@uky.edu
 \linebreak
Phone:  (859)-257-1397
 \linebreak
 
\textbf{Co-authors:} 
R.\ Kraft (Smithsonian Astrophysical Observatory), E.\ Roediger (University of Hull), P.\ Nulsen (Smithsonian Astrophysical Observatory), A.\ Sheardown (University of Hull), 
T.\ Fish (University of Hull), J.\ ZuHone (Smithsonian Astrophysical Observatory), E.\ Churazo (Max Planck Institute for Astrophysics), W.\ Forman (Smithsonian Astrophysical Observatory), C.\ Jones (Smithsonian Astrophysical Observatory), J.\ Irwin (University of Alabama), S.\ Randall (Smithsonian Astrophysical Observatory)
%R.\ Kraft\footnote{Center for Astrophysics | Harvard \& Smithsonian, 60 Garden Street, Cambridge, MA 02138, USA},
%{E.\ Roediger}\footnote{E.A. Milne Centre for Astrophysics, Department of Physics and Mathematics, University of Hull, Hull, HU6 7RX, United Kingdom},
%{P.\ Nulsen}$^{1}$,
%{A.\ Sheardown}$^{2}$,
%T.\ Fish$^{2}$,
%{J.\ ZuHone}$^{1}$,
%{E.\ Churazov}\footnote{Max Planck Institute for Astrophysics, Karl-Schwarzschild-Str. 1, 85741, Garching, Germany},
%{W.\ Forman}$^{1}$,
%{C.\ Jones}$^{1}$,
%{J.\ Irwin}\footnote{Department of Physics and Astronomy, University of Alabama, Box 870324, Tuscaloosa, AL 35487, USA},
%{S.\ Randall}$^{1}$
\linebreak
\justify
\textbf{Abstract:}

Clusters of galaxies are the most massive virialized structures in the Universe.
%Most of the hot baryons in the Universe are in the form of intracluster plasma, emitting X-rays through thermal bremsstrahlung and atomic lines. 
The microphysical properties of the intracluster plasma can affect dynamical processes over an enormous range: from the feedback of active galactic nuclei to shock acceleration in merging clusters. All the major cosmological simulations assume the astrophysical plasma to be inviscid. It is critical to measure microphysical properties of the intracluster plasma to truly understand the physical processes that drive the cosmic evolution. Tremendous progress has been made by comparing high spatial resolution X-ray images to (magneto-)hydrodynamic simulations. Future X-ray missions with calorimeters promise a direct measurement of transport coefficients and gas motions, providing a more realistic benchmark for cosmological simulations.  
% With {\sl Chandra} and {\sl XMM-Newton} observations and (magneto-)hydrodynamic simulations,  tremendous progress has been made in understanding how clusters are assembled and how energy is transported in the ICM. 

\pagebreak

%\begin{multicols}{2}     
\section{The intracluster medium as a magnetized plasma}

\begin{figure}
\floatbox[{\capbeside\thisfloatsetup{capbesideposition={right,top},capbesidewidth=6.3cm}}]{figure}[\FBwidth]
{\caption{Large-scale Gas Distribution of IllustrisTNG: the Illustris follow-up simulation (\url{www.tng-project.org}). All the current major cosmological simulations assume viscosity to be zero. It is critical to measure the viscosity of the gas in galactic systems and large scale structures to truly understand the physical processes that drive the cosmic evolution.}\label{fig:tng}}
 { \includegraphics[width=0.5\textwidth]{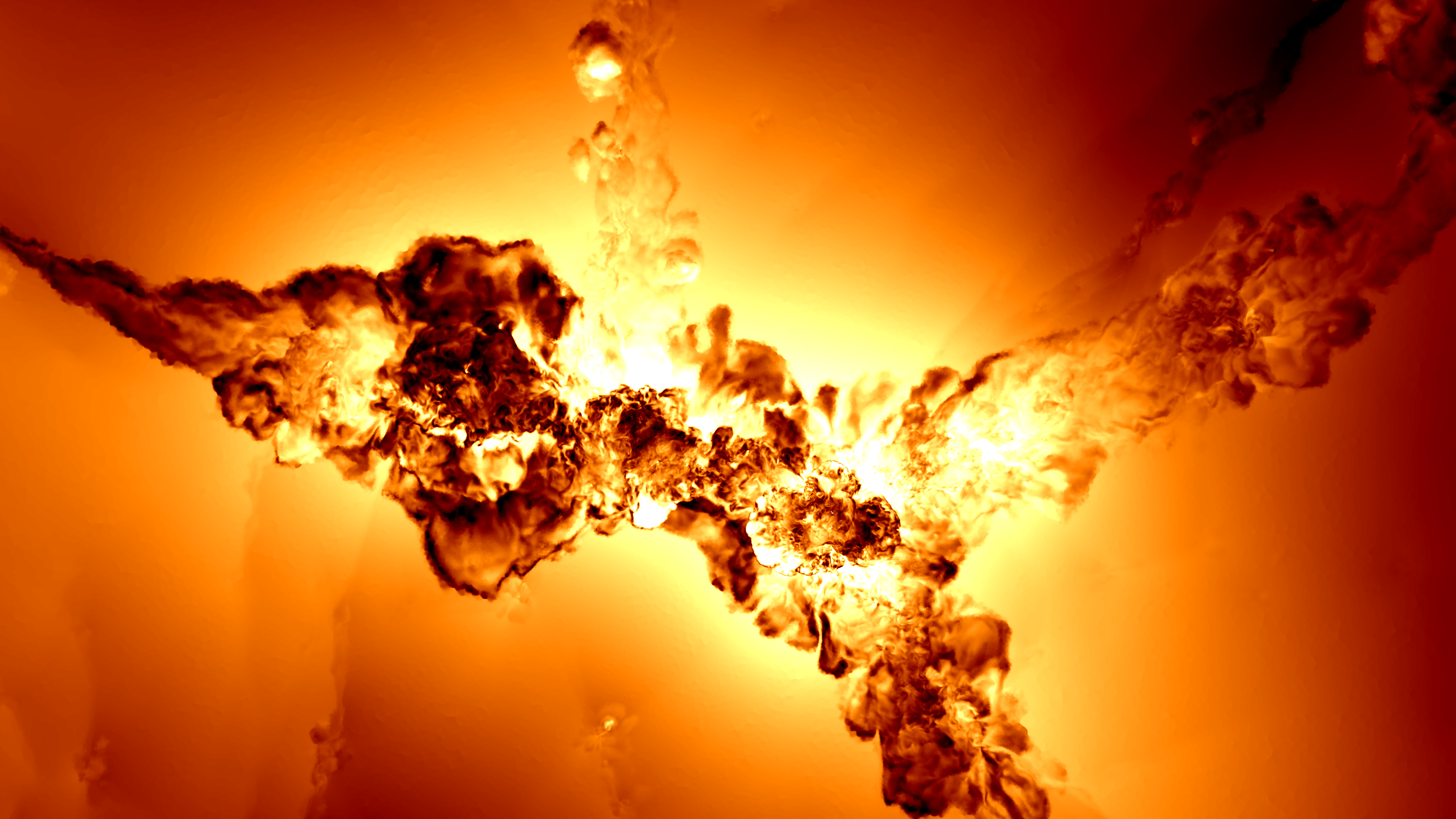}}
 \vspace{-0.2cm}
%\hspace{-0.4cm}{ \includegraphics[width=0.6\textwidth]{v3}}
\end{figure}

%Clusters of galaxies are the most massive ($10^{14}-10^{15}$\,M$_{\odot}$) virialized structures in the Universe. 
%As the tracers of massive halos over cosmic time, galaxy clusters have been appreciated as cosmology probes (Allen et al.\ 2011).
%They are the end result of the hierarchical structure formation and contain most of the bulk of all dark matter, hot baryons, and metallicity. 
The space between cluster member galaxies is filled with hot and diffuse ionized plasma that emits X-rays through thermal bremsstrahlung and atomic lines. 
%Our knowledge of fluid mechanics 
%that we learn from oceanography, meteorology, and space science 
%has been used to understand the dynamics of the intracluster medium (ICM).
% despite the fact that galaxy clusters are on a scale of $\sim$Mpc. 
%Meanwhile, the ICM can be used to study hydrodynamics under extreme conditions --- its diffuseness, high temperature, and the variety of physical scales cannot be created in terrestrial laboratories.  
%It has been challenging to measure cluster mass accurately. 
%The most practical way of measuring cluster mass for cosmological study is based on the assumption that the cluster gas is in hydrostatic equilibrium in their combined gravitational potential. However, the ICM is always in a dynamically active state due to its continuing growth. 
%Galaxy clusters grow via mergers of less massive subclusters.
%We can trace the motion of infalling subclusters by solving Euler equations and study features of bow shocks ahead of infalling objects by applying Rankine–Hugoniot conditions. These methods have been used to decipher the assembly history of galaxy clusters and (re)energizing processes in the ICM (e.g., Vikhlinin et al.\ 2001a,b; Su et al.\ 2017a,b;  Markevitch et al.\ 2002; Botteon et al.\
%2018). The validity of these approaches are limited by projection effects, by the non-spherical shape of subclusters, and the fact that the ICM is not an ideal fluid flow (Zhang et al.\ 2019).  
 %\noindent{-Motivation: why it is important to study hydrodynamics in galaxy clusters?}
%The  intracluster medium (ICM) is a magnetized plasma, and 
We often treat it as a fluid with transport coefficients, where the particle-magnetic field interactions and plasma instabilities determine the effective transport coefficients. 
It is critical to understand the transport processes in the ICM on a variety of physical scales: 
microscopic plasma physics impacts many macroscopic processes, e.g., the dissipation and redistribution of the kinetic energy released by mergers or AGN outbursts (Fabian et al.\ 2005), the cooling and heating in cluster cores, and gas stripping from galaxies (Nulsen 1982). 
%To date, a number of microphysical properties of the ICM remain ill-constrained,  
Furthermore, large cosmological simulation is our primary way to understand the baryonic processes of the Universe. 
However, all the current major cosmological simulations assume the interstellar medium, intracluster medium, and large scale gas to be inviscid (Figure~\ref{fig:tng}) i.e., illustris project (Vogelsberger et al.\ 2014), the EAGLE project (Schaye et al.\ 2014) (although the numerical viscosity is non-zero and depends on the resolution of the simulation). To fully understand 
feedback, galaxy evolution, and structure formation etc., it is crucial to constrain the microphysical properties of the astrophysical plasma including viscosity, thermal conductivity, the strength and topology of the magnetic field, and small-scale turbulence.

X-ray observations have been used to probe the physical conditions in the ICM. For example, turbulent velocity has been determined indirectly via resonance scattering and surface brightness fluctuations (Ogorzalek et al.\ 2017; Gu et al.\ 2018; Zhuravleva et al.\ 2015). The microcalorimeter on board Hitomi has measured the level of turbulence directly using line broadening (Hitomi collaboration 2016). 
The {\sl Chandra} X-ray Observatory (half arcsec spatial resolution) has revealed the ubiquity of ``cold fronts" in the ICM (see Markevitch \& Vikhlinin 2007 for a review), which can be used to constrain the viscosity and magnetic field. Cold fronts are sharp
interfaces between cooler, denser, hence brighter, gas and
hotter, lower density, hence fainter, medium, where the pressure is
continuous.  They result from merger activity, either created directly
by the infall of a low entropy subcluster (merging cold front), or
from gas motions induced by the gravity of an infalling subcluster
(sloshing cold fronts). 
For purely
hydrodynamic interactions, Kelvin-Helmholtz Instability (KHI) is expected to
develop at shearing interfaces.  However, magnetic fields or viscosity
can suppress the KHI (Chandrasekhar 1961; Lamb 1932).  State of the art simulations have
demonstrated that cold fronts appear smoother in the presence of
either magnetic field or viscosity at the levels expected in the ICM
(see ZuHone \& Roediger 2016 for a review).

%\begin{figure}
%\floatbox[{\capbeside\thisfloatsetup{capbesideposition={right,top},capbesidewidth=5cm}}]{figure}[\FBwidth]
%{\caption{ The existing (left) and suppressed (right) KHI found in our simulations of typical cluster cold fronts under different conditions. {\it top-left}: no magnetic field. {\it top-right}: strong magnetic field. {\it bottom-left}: low viscosity. {\it bottom-right}: high viscosity.}\label{fig:test}}
 %{ \includegraphics[width=0.5\textwidth]{Figure1}}
%\hspace{-0.4cm}{ \includegraphics[width=0.6\textwidth]{v3}}
%\end{figure}

\begin{figure}
\floatbox[{\capbeside\thisfloatsetup{capbesideposition={right,top},capbesidewidth=5.8cm}}]{figure}[\FBwidth]
{\caption{ {\bf Top:} The {\sl Chandra} X-ray image of Abell~3667 (left) was compared with the simulated gas stripped galaxy for an inviscid atmosphere (right). The figure is taken from Ichinohe et al.\ (2017). {\bf Middle:} The {\sl Chandra} X-ray image of the Perseus cluster (left) was compared with the simulated cluster with an initial magnetic field of $\beta=200$ (right). The figure is taken from Walker et al.\ (2017). {\bf Bottom}: The temperature map of the infalling galaxy NGC 1404 obtained with deep Chandra observations (left, Su et al.\ 2017a) was compared with the temperature map produced by the numerical simulation specifically tailored to the case of NGC 1404 falling into the Fornax Cluster (right, Sheardown et al.\ 2018).}\label{fig:test}}
 { \includegraphics[width=0.5\textwidth]{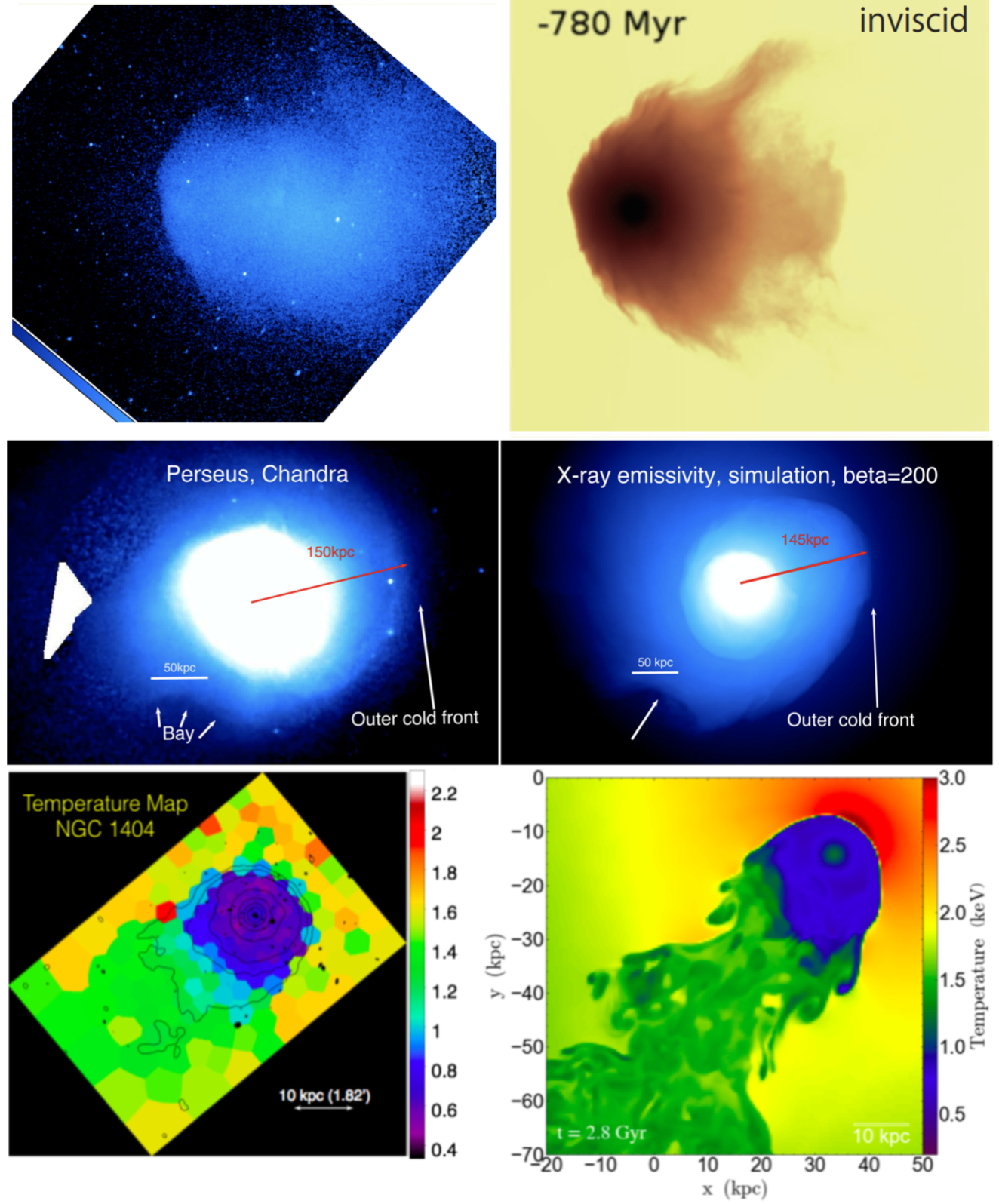}}
\vspace{-0.2cm}
%{ \includegraphics[width=0.6\textwidth]{v3}}
\end{figure}

Recent deep {\sl Chandra} observations have been dedicated to revisiting cold fronts identified by previous observations, leading to a deeper understanding of the microphysics of the ICM. 
Multiple-edge structures in X-ray surface brightness have been identified at the cold fronts in a growing number of systems (Werner et al.\ 2016; Su et al.\ 2017a; Ichinohe et al.\ 2017; Ichinohe et al.\ 2018). 
%For example, recent azimuthal studies of the merging cold fronts in Abell 3667 and NGC~1404 reveals that cold fronts edges in the surface brightness profiles across the cold front (Ichinohe et al.\ 2017; Su et al.\ 2017a). 
These features are consistent with the presence of KHI eddies and an inviscid ICM (Figure~\ref{fig:test}-top).
% $<5\%$ of the isotropic Spitzer-like viscosity can be derived.
Walker et al.\ (2017) studied the sloshing cold fronts in a number of clusters (Perseus, Centaurus and Abell 1795) and identified concave `bay' substructures from X-ray and radio imaging. These features resemble the giant KHI rolls expected when the ratio of thermal to magnetic pressure is
$\beta=200$ (Figure~\ref{fig:test}-middle).

In addition to cold fronts, stripped tails are another product of ongoing infall in galaxy clusters and can be used to constrain micro physics of the ICM (Roediger et al.\ 2015a,b). For example, a temperature gradient was detected in the stripped tail in NGC~1404, indicating a well mixed plasma of a low viscosity (Su et al.\ 2017a; Sheardown et al.\ 2018). The temperature map derived with the deep {\sl Chandra} observations is remarkably similar to that derived from an inviscid simulation specifically tailored to the infall of NGC~1404 (Figure~\ref{fig:test}-bottom). 
Results from deep X-ray observations of the nearby early-type galaxy NGC~4552 were compared with the viscous and inviscid hydrodynamic simulations specifically tailored to the stripping of this galaxy (Roediger et al.\ 2015a,b; Kraft et al.\ 2017). Inviscid stripping was favored by the study. Based on the survival of the stripped low-entropy infalling group in the hot cluster Abell~2142, Eckert et al.\ (2014, 2017) find thermal conduction to be strongly suppressed.
Note that the properties of cold fronts and stripped tails also depend on the merger history (Su et al.\ 2017d; Kraft et al.\ 2017). It is important to understand the entire dynamic state of the system before any conclusions can be drawn from the immediate observables.  
      
\section{Measuring gas motions in galaxy clusters}

The ICM is always in a dynamically active state. Forming at the knots of the cosmic filaments, galaxy clusters grow continually via mergers of subclusters and accretions of galaxies. In addition, AGN periodically release mechanical energy at cluster centers. Both effects are expected to cause shocks, bulk motions, and turbulence in the ICM over a large span of physical scales. To date, the most practical way of probing the ICM velocity is based on the Rankine-Hugoniot jump conditions: the gas properties on both sides of a shock wave can be used to infer the infalling speed of a substructure (e.g, Su et al. 2016; Vikhlinin et al. 2001; Markevitch et al. 2002; Zhang et al.\ 2019). For infalling substructures, Su et al.\ (2017b) has developed an analytical method combining the jump conditions and the Bernoulli equation, which allows us to derive a general speed, the complete velocity field, and the distance of the substructure from the pressure distribution. 
%The authors have applied this method in the deep {\sl Chandra} observations of the nearby X-ray bright elliptical galaxy NGC~1404 falling through the ICM of the Fornax Cluster. The best-fit results imply that NGC~1404 is falling at a Mach number of 1.3 at an inclination angle of $33^{\circ}$ and at the same distance as the Fornax center. These results have been reproduced in a hydrodynamics plus N+body simulations specifically tailored to the merger of Fornax and NGC~1404 (Sheardown et al.\ 2018). 
The motion of gas sloshing can also be determined via  the pressure distributio. Ueda et al.\ (2018) analyzed the cool core cluster RX J1347.5-1145. They found that while the residual X-ray image derived from {\sl Chandra} shows a clear spiral pattern characteristic of gas sloshing, no significant variation is shown in the Sunyaev-Zel'dovich effect image derived from {\sl ALMA}. This study has confirmed the subsonic nature of sloshing gas predicted by simulations (e.g., Roediger et al.\ 2011; ZuHone et al.\ 2010). 

The line-of-sight gas motion can be measured directly from line centroids based on the Doppler effect. The line centroids measurements have been performed in a number of galaxy clusters with CCDs (Dupke \& Bregman 2006; Ota et al.\ 2007; Tamura et al.\ 2011; Ueda et al.\ 2019). However, these results are mostly upper limits or of low significance due to the poor spectral resolution of CCDs ($\Delta E\sim150 {\rm eV}$). 
In contrast, the microcalorimeter on board {\sl Hitomi} provides a spectral resolution of 5\,eV. Based on the FeXXV K$\alpha$ line at 6.7\,keV, {\sl Hitomi} accurately measures a bulk motion of 150\,km/s for the gas at the center of the Perseus Cluster. Although this is the only ICM measurement made by {\sl Hitomi} due to its untimely loss, it has demonstrated that  calorimeter science can revolutionize our view of the gas dynamics of the ICM.     

%\begin{figure}
%\floatbox[{\capbeside\thisfloatsetup{capbesideposition={right,top},capbesidewidth=7.5cm}}]{figure}[\FBwidth]
%{\caption{ Observation and simulation investigation of the Fornax Cluster, the nearest cluster in the southern sky (Su et al. 2017a,b,c,d). {\bf top-left}: XMM-Newton mosaic image of the Fornax Cluster using 17 EPIC pointings in units of cts/s/deg$^2$. A number of sloshing cold fronts, triggered by off-axis mergers, can be identified. {\bf top-right}: The density map produced by our numerical simulation specifically tailored to the case of NGC 1404 falling into the Fornax Cluster. {\bf bottom-left}: The temperature map of the infalling galaxy NGC 1404 in units of keV obtained with deep Chandra observations. Black contour: Chandra X-ray image. {\bf bottom-right}: The temperature map produced by our numerical simulation specifically tailored to the case of NGC 1404 falling into the Fornax Cluster (zoomed in on “NGC 1404}\label{fig:test}}
% { \includegraphics[width=0.5\textwidth]{fornax}}
%\hspace{-0.4cm}{ \includegraphics[width=0.6\textwidth]{v3}}
%\end{figure}

\section{What can we learn with future X-ray telescopes?}

\begin{figure}[h]
\centering
 %      \vspace{-0.3 cm}
   \includegraphics[width=0.49\textwidth]{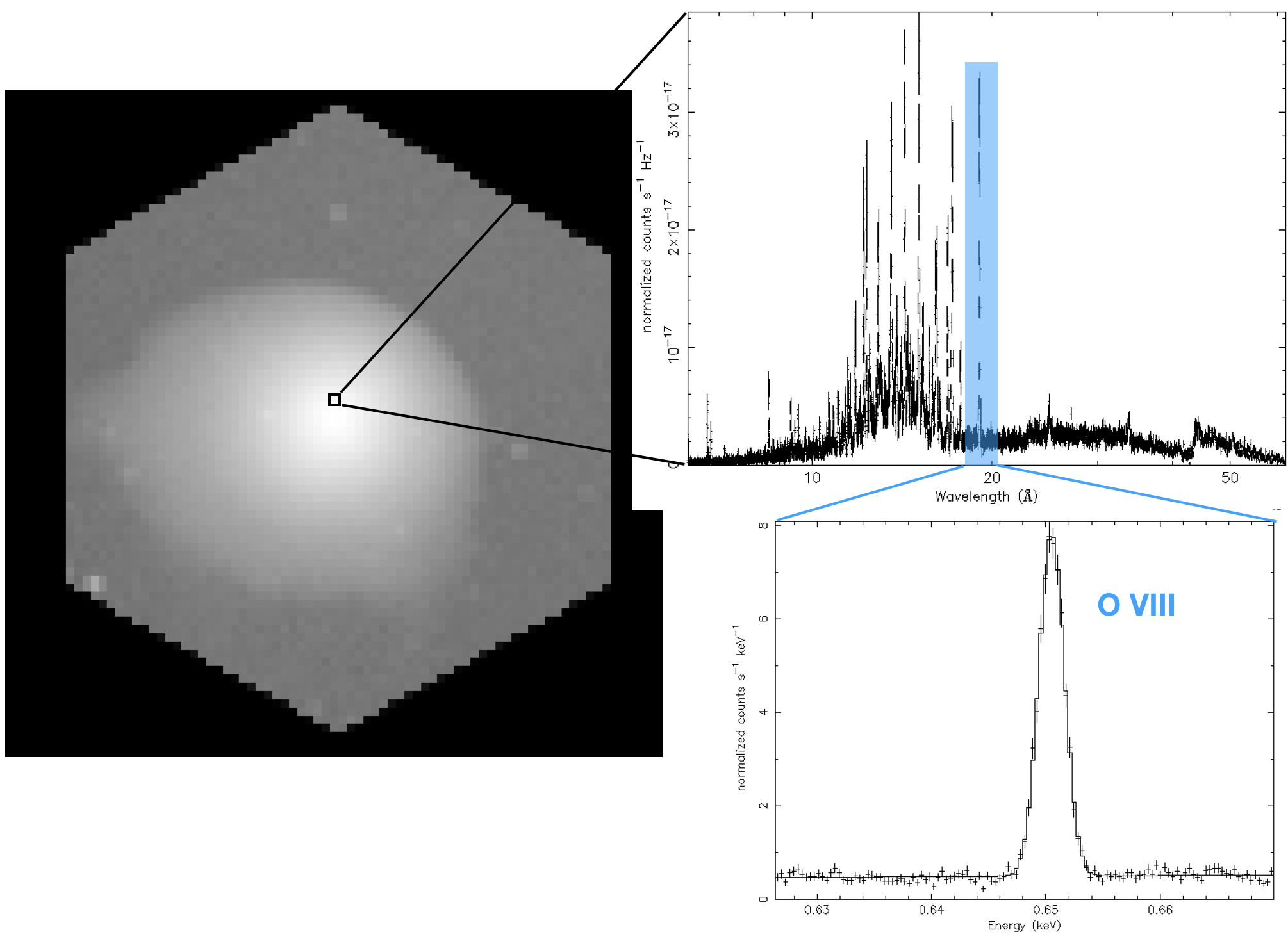}
     \includegraphics[width=0.42\textwidth]{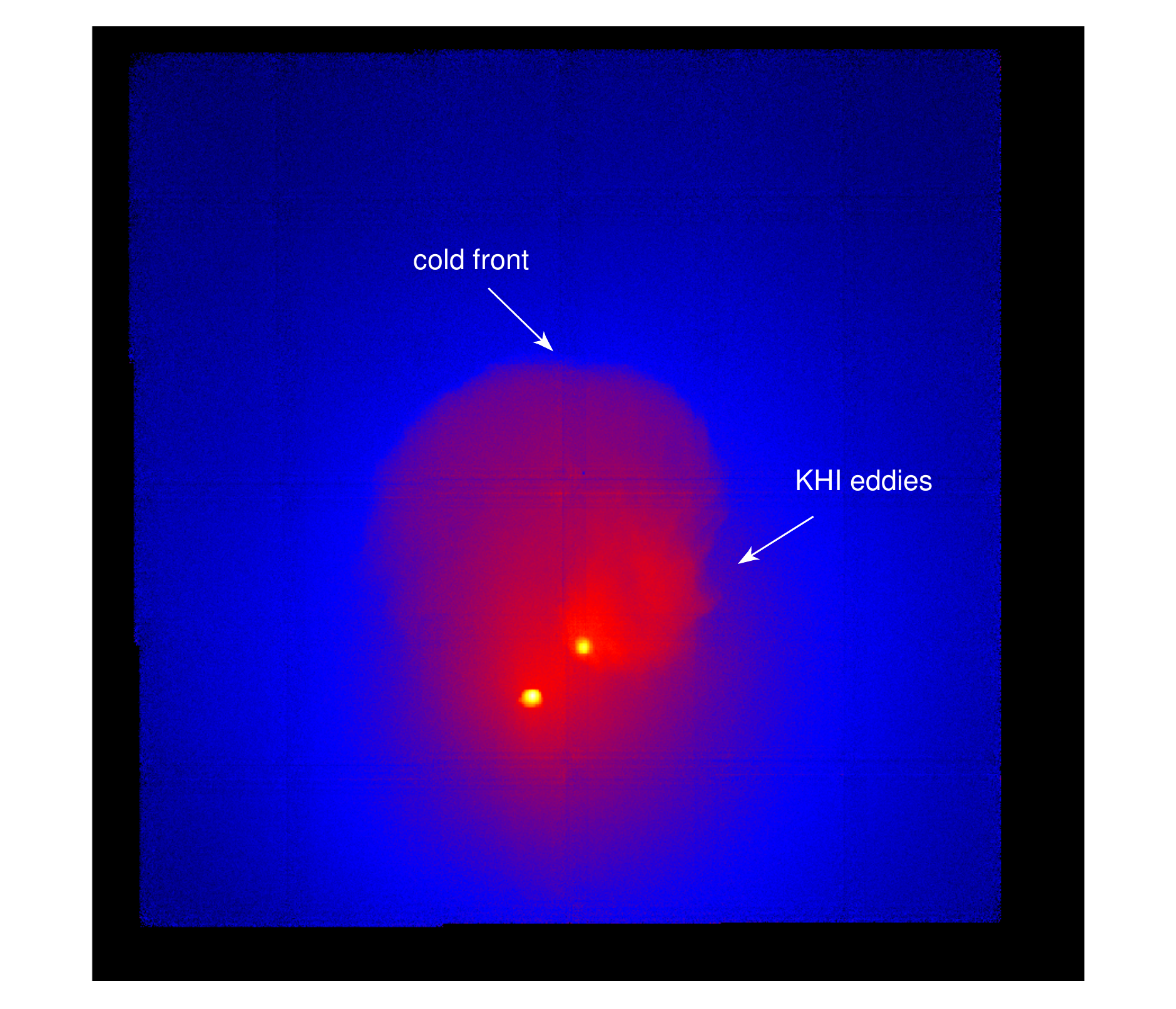}
       \caption{\footnotesize \label{fig:6}  {\bf Left:} Simulated 200\,ksec X-IFU observation of NGC~1404 in the 0.5--2.0 keV energy band. The O VIII line is relatively isolated and sufficiently bright, which can be used to constrain gas dynamics in low mass clusters and galaxies.  {\bf Right:} Simulated WFI 3X3 30\,ksec mosaic observations of Fornax. Subtle substructures and the cold front at large radii can be detected.}
       \vspace{-0.4 cm}
       \end{figure}

The biggest change promised by future missions is to be able to measure the kinematics of the intracluster medium directly with calorimeters at high spatial resolution. Thanks to the hierarchical formation process, massive clusters are the rare breed in the Universe. Galaxy groups and low mass clusters are much more common, for which the measurement of FeXXV K$\alpha$ line at 6.7\,KeV is not available. Instead, the centroid and broadening of the O VIII line at $18.967\mathring{A}$ (0.654\,keV), which is
bright and isolated for the hot atmospheres of low mass clusters and
galaxies, can be used to map the bulk speed and turbulence of the
gas. 
Using the FLASH simulation (Sheardown et al.\ 2018), sixte 1.3.6 and SOXS 2.2.0,
we simulated a 200\,ksec observation of a nearby X-ray bright galaxy like NGC~1404 using an instrument with an effective area of 1.4\,$m^2$ and with a superb spectral resolution, similar to the X-ray Integral Field Unit (X-IFU) on board Athena (Barret et al.\ 2018). A simulated image and an example spectrum is shown in Figure~\ref{fig:6}-left, demonstrating its 2.5 eV spectral resolution with $5^{\prime\prime}$ pixels.

%Using the FLASH simulation (Sheardown et al.\ 2018), sixte 1.3.6 and SOXS 2.2.0., Our observation is focused on NGC~1404. Unlike more massive galaxy clusters such as Perseus and Coma, the measurement of FeXXV K$\alpha$ line at 6.7\,KeV is not available for low mass clusters. 

We have also simulated a $3\times3$ mosaic of 30\,ksec observations for an instrument with a large field of view and low and stable background resembling the wide field X-ray imager (WFI) on board Athena as shown in Figure~\ref{fig:6}-right. We can capture the entire ICM of a cluster like Fornax out to (and even beyond) the virial radius and detect fainter sloshing cold fronts at larger radii, which is not feasible with existing missions for a reasonable exposure time. We have selected regions of interest to directly measure their gas motions and turbulence with X-IFU simulations
We can further identify uneven edges at the cold front and turbulent regions in the path of the infalling object. The details of these faint structures match the resolution of the simulations, which will transform our understanding of the microphysics over a large span of radii the ICM.  
Gas distribution at large radii will also pin down the initial conditions for the simulation, maximizing the science return from the tailored simulations. 

\section{Concluding Remarks}
The vast bulk of the hot baryons in the Universe is in the form of the intracluster medium, a hot diffuse plasma emitting X-ray via thermal bremsstrahlung and atomic lines.
 With {\sl Chandra} and {\sl XMM-Newton} observations and (magneto-)hydrodynamic simulations, tremendous progress has been made in understanding how clusters are assembled and how energy is transported in the ICM. 
Future missions such as XARM, Athena, and Lynx will be equipped with calorimeters, allowing us to measure gas motions and turbulence directly and to put quantitative constraints on the transport coefficients in their ICM. 
By the end of the 2020's, hundreds of thousands galaxy clusters will be detected by various cluster surveys such as eROSITA (Pillepich et al.\ 2018).  
Most of them are expected to be low mass clusters. The O VIII line is relatively isolated and sufficiently bright, which can be used to constrain gas dynamics in low mass clusters and galaxies.
Through a connection between the micro and macro scale astrophysics in the ICM, our knowledge of hydrodynamics in galaxy clusters will be revolutionized.  
%\end{multicols}

\pagebreak
%\textbf{References}

\end{document}